\documentclass[structabstract]{aa}  
\usepackage{graphicx}
\usepackage{txfonts}
\usepackage{amsfonts}
\usepackage{subfigure}
\usepackage{amssymb,amsfonts}
\usepackage{natbib}
\bibpunct{(}{)}{;}{a}{}{,}
\begin{document}
\title{First fringes with an integrated-optics beam combiner at 10\,$\mu$m}

\subtitle{A new step towards instrument miniaturization for mid-infrared 
interferometry}

\author{L. Labadie\inst{1,4,5}, G. Mart\'in\inst{2}, N. C. Anheier\inst{3}, B. Arezki\inst{2}, H. A. Qiao\inst{3}, B. Bernacki\inst{3}, P. Kern\inst{2}
}

\institute{
Instituto de Astrofisica de Canarias, C/ Via Lactea s/n, La Laguna, Tenerife E-38200, Spain \\ \email{labadie@iac.es}
\and
UJF-Grenoble 1/CNRS-INSU, Institut de Plan\'etologie et d'Astrophysique de Grenoble (IPAG) UMR 5274, Grenoble, F-38041, France
\and
Pacific Northwest National Laboratory, 902 Battelle Boulevard, P.O. Box 999, Richland, Washington 99352, U.S.A
\and
Departamento de Astrofisica, Universidad de La Laguna, 38205 La Laguna, Tenerife, Islas Canarias, Spain
\and
I. Physikalisches Institut, Universit\"at zu K\"oln, Z\"ulpicher Str. 77, 50937 K\"oln, Germany
}

\date{Received September 15, 1996; accepted March 16, 1997}
 
  \abstract
   {Observations at milliarcsecond-resolution scales and high dynamic range hold a central place in the exploration of distant planetary systems to achieve, for instance, the spectroscopic characterization of exo-Earths or the detailed mapping of their protoplanetary discs birthplace. Multi-aperture infrared interferometry, either from the ground or from space, is a very promising technique to tackle these goals. However, significant technical efforts still need to be undertaken to achieve a simplification of these instruments if we want to recombine the light from a large number of telescopes. Integrated-optics concepts appear as an alternative to the current conventional designs, especially if their use can be extended to a higher number of astronomical bands.
}
   {This article reports, for the first time to our knowledge, the experimental demonstration of the feasibility of an integrated-optics approach to mid-infrared beam combination 
for single-mode stellar interferometry.}
   {We have fabricated a 2-telescope beam combiner prototype integrated on a substrate of chalcogenide glass, a material transparent from $\sim$1\,$\mu$m to $\sim$14\,$\mu$m. We have developed laboratory tools to characterize in the mid-infrared the modal properties and the interferometric capabilities of our device.}
   {We obtain interferometric fringes at 10\,$\mu$m and measure a mean contrast $V$=0.981$\pm$0.001 with high repeatability over one week and high stability over a time-period of $\sim$5\,h. We show experimentally -- as well as on the basis of modeling considerations -- that the component has a single-mode behavior at this wavelength, which is essential to achieve high-accuracy interferometry. From previous studies, the propagation losses are estimated to 0.5\,dB/cm for this type of component. We also discuss possible issues that may impact the interferometric contrast.}
   {The IO beam combiner performs well at the tested wavelength. We also anticipate the requirement of a better matching between the numerical apertures of the component and the (de)coupling optics to optimize the total throughput. The next step foreseen is the achievement of wide-band interferograms. 
}

\keywords{Instrumentation: high angular resolution, interferometers -- Methods: laboratory -- Techniques: interferometric}

\authorrunning{L. Labadie et al.}
\titlerunning{First fringes with an integrated-optics beam combiner at 10\,$\mu$m}

\maketitle
%

\section{Introduction}\label{Intro}

Very recent results have demonstrated the potential of optical/IR interferometers to produce the first infrared astrophysical images of complex morphologies at the milli-arcsecond scale. These images have revealed unprecedented details on the highly distorted photosphere of the fast rotator Altair \citep{Monnier2007}, they have produced the first direct view on the inner regions of a young circumstellar disk \citep{Renard2010}, and have permitted to measure the gas velocity map in the very close surrounding of a supergiant \citep{Millour2011}. Such important breakthroughs would have not been possible without the help of imaging interferometers. 
Key to this technique is on the astronomers ability to simultaneously and coherently combine the beams from three or more telescopes. This permits the implementation of closure-phase techniques to obtain the interferometric visibilities amplitude and phase and allows reconstruction of a complex astrophysical image. 
Although other techniques exist to retrieve the necessary phase information, increasing the number of sub-apertures and using Earth-rotation synthesis also drastically improves the UV plane coverage. 
This provides nearly snapshot reconstructed images, exhibiting high-fidelity and model-independent views of the object. 
However, the corollary of multi-beam combination is a significant increase in the optical complexity of the interferometer coupled to the need for high instrumental stability (mechanical and thermal) and for a good level of control of the shape of the incoming wavefronts. 
In the last decade, instrumental research programs have demonstrated the feasibility of {\it single-mode} integrated optics (IO) components to enable both multi-aperture and high accuracy interferometric instruments using centimeter-scale devices (\cite{Malbet99} and the following papers series). 
IO-based solutions have proven their astronomical potential with IOTA/IONIC \citep{Berger2001,Monnier2004,Kraus2005} and VLTI/PIONER, putting them eventually at the heart of the future interferometric instrument GRAVITY at the VLTI. 
To date, this elegant solution and hence the resulting science has been limited to the near-infrared domain (J,~H,~K bands), yet there is an unquestionable astrophysical interest to extend this approach to the mid-infrared range, beyond $\sim$3\,$\mu$m. 
This is for instance a key spectral range where to study objects with temperatures of $\sim$100--600\,K (e.g. the planet forming regions around young solar-type stars, the location of the snow line \citep{Sasselov2000}, debris disks morphologies as remnant of the planet formation process). Chemical bio-markers like CO$_{2}$ or H$_{\rm 2}$O could be detected spectroscopically between 5 and 20\,$\mu$m range to characterize the atmosphere of exoplanets in the habitable zone of solar-type stars \citep{Cockell2009}. 
In this respect, a space-based mid-infrared and multi-aperture {\it nuller} able to produce a deep extinction of the stellar light \citep{Angel1997,Mennesson1997} would benefit as well from the compactness and stability of a IO beam-combiner. 
These different aspects have motivated a long-term work in the community in order to extend the IO instrumental solution beyond 2\,$\mu$m \citep{Mennesson1999,Laurent2002,Wehmeier2004,Labadie2006,Labadie2007,Hsiao2009}. 
Because of the relative technological gap historically existing with the more mature near-infrared solutions, the mid-infrared extension has required a 
concerted R\&D research effort between astronomers and photonics experts.\\
\indent In this study, we demonstrate for the first time the fabrication and the operation of a two-telescope IO single-mode beam combiner for the mid-infrared spectral range around $\lambda$=10\,$\mu$m. 
Chalcogenide glass materials are used for fabrication of this IO device, primarily due to the excellent infrared transparency in the 1--14\,$\mu$m range 
and the photo-induced modification properties, i.e. the permanent or reversible increase of a glass refractive index induced by radiation (light, laser irradiation, etc...). The present work focuses primarily on the achievable interferometric contrasts and on the single-mode characterization of this new component.

\begin{figure}[b]
\centering
\includegraphics[width=4.5cm]{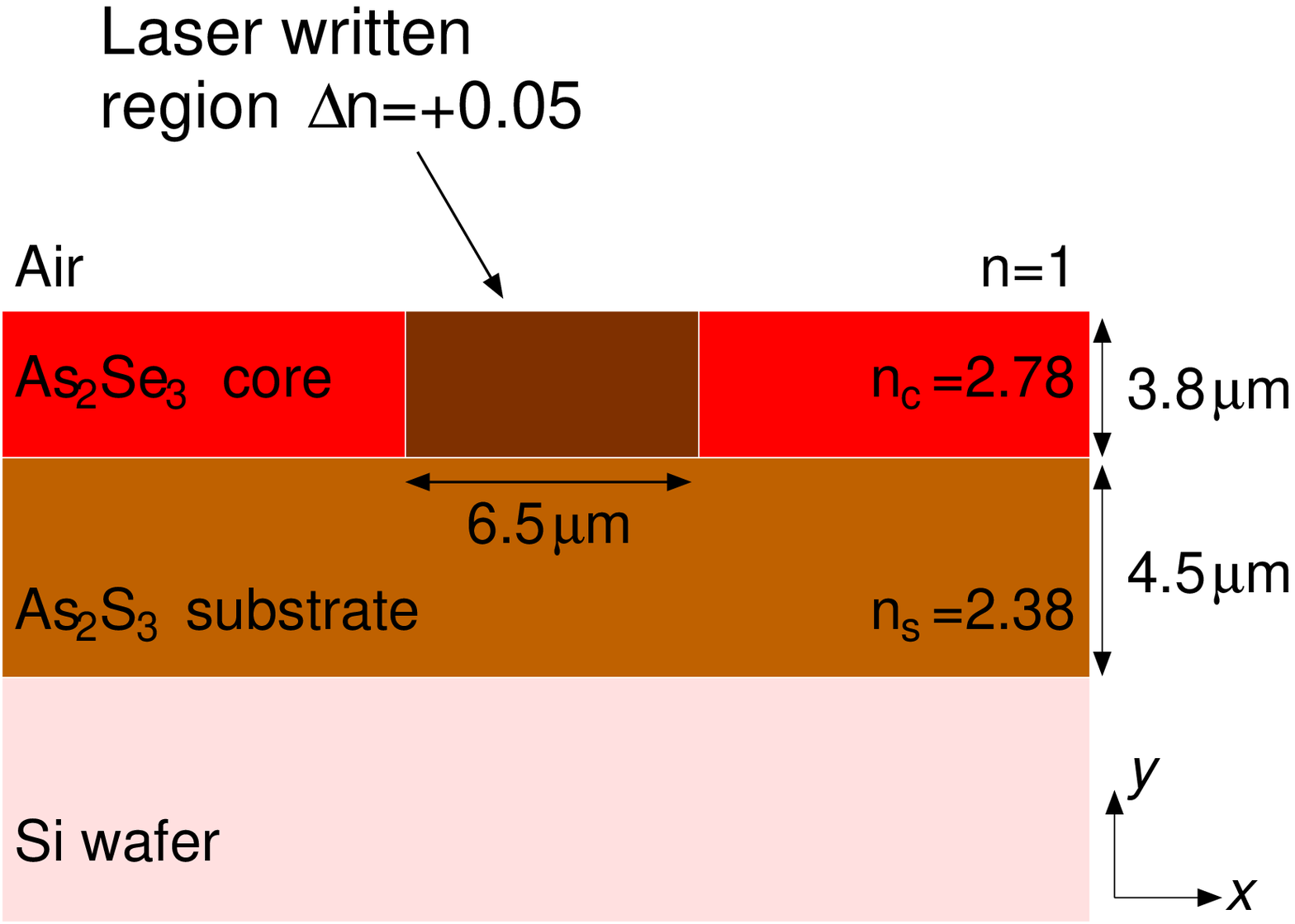}
\hspace{0.1cm}
\raisebox{0.0cm}{\includegraphics[width=4.0cm]{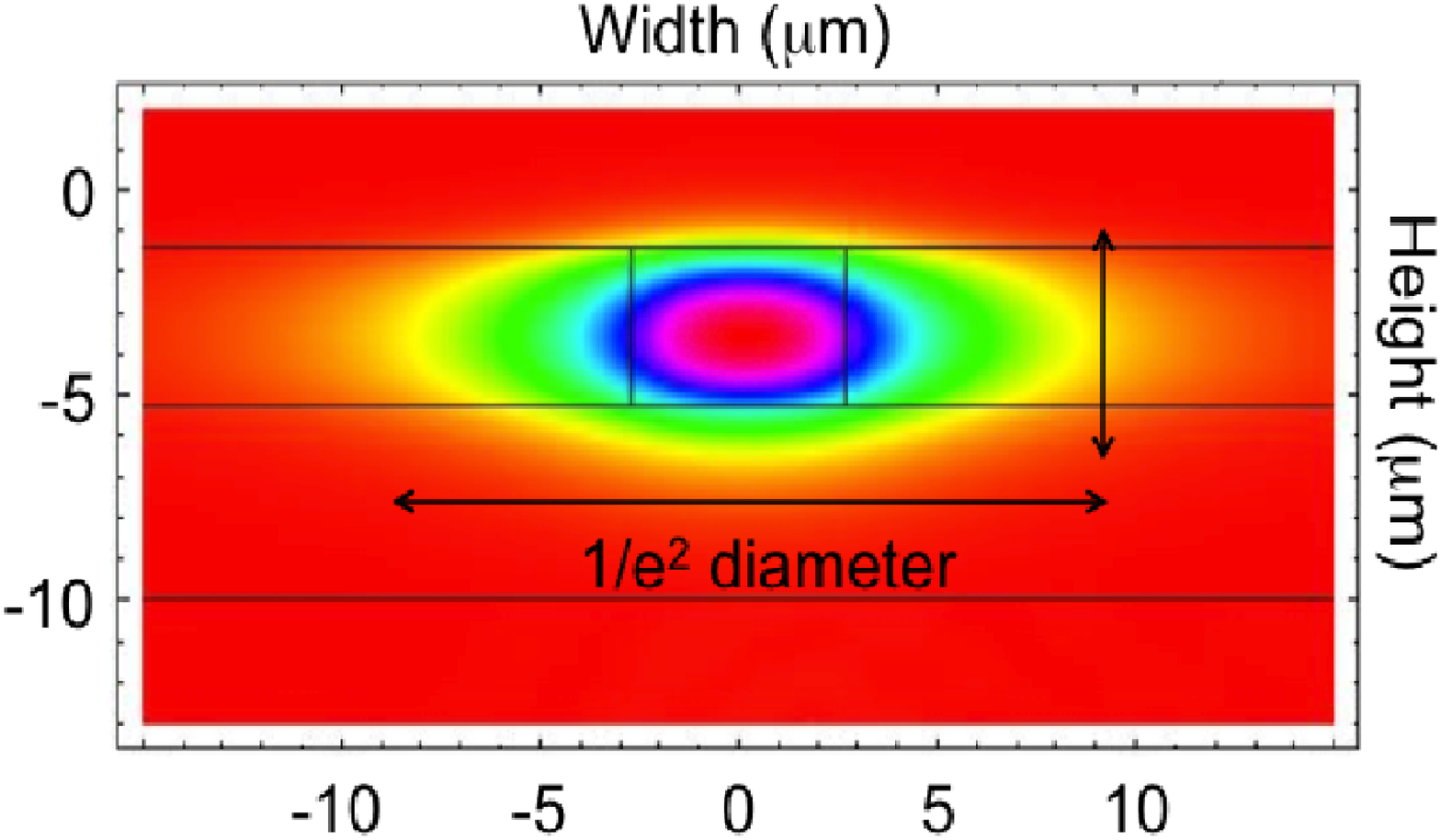}} 
\caption{{\it Left:} Cross-section view of the Y-junction geometry. The high-index strip waveguide is bounded by As$_{\rm 2}$Se$_{\rm 3}$, As$_{\rm 2}$S$_{\rm 3}$ and air. Examples of refractive indices at 6, 8, 10 and 12\,$\mu$m used in this study are:  
$n$=2.788, 2.783, 2.777, 2.774 for As$_{\rm 2}$Se$_{\rm 3}$; 
$n$=2.403, 2.394, 2.381, 2.364 for As$_{\rm 2}$S$_{\rm 3}$. 
{\it Right:} theoretical spatial distribution of the electric field for the TE$_{\rm 0,0}$ fundamental mode superimposed to the waveguide geometry. The vertical scale ranges from high (violet) to low (yellow) intensity. }\label{geometry}
\end{figure}

\section{Component design and manufacturing}

\subsection{Dimensioning and modal behavior}\label{modal-behav}

The opto-geometrical parameters of the Y-junction\footnote{The Y-junction is a photonics device which can be used either as a beam-splitting device (1 input, 2 outputs) or as a beam-combiner device (2 inputs, 1 input).} are chosen to constrain the {\it single-mode} behavior in the 10\,$\mu$m spectral range in order to benefit from the necessary spatial filtering of the combined wavefronts. 
Since a vast bibliography is existing on the advantage of spatial filtering by single-mode waveguides for stellar interferometry and the corresponding theory (see \cite{Foresto1997,Mennesson2002} and citing papers), we do not treat this aspect here. 
In Fig.~\ref{geometry}--left we illustrate the geometry adopted in this study, which refers to an asymmetric strip embedded waveguide configuration. The electric field confinement in the $y$-direction is ensured by the high-index core layer of As$_{\rm 2}$Se$_{\rm 3}$ surrounded by a low-index substrate layer of As$_{\rm 2}$S$_{\rm 3}$ and air ($n$=1). The confinement in the $x$-direction is ensured by an increase of the refractive index of the core layer over a well-defined region of the planar waveguide. The parameters considered to describe theoretically the modal behavior of the channel waveguide are the thickness $d$=3.8\,$\mu$m and width $w$=6.5\,$\mu$m of the strip, a local increase of the refractive index $\Delta n$=+0.05. The refractive index value as a function of the wavelength for As$_{\rm 2}$Se$_{\rm 3}$ and As$_{\rm 2}$S$_{\rm 3}$ are taken from the literature (\cite{Savage1965,Palik1985,Klocek1991}; see examples in the caption of Fig.~\ref{geometry}). We compute the theoretical bi-mode/single-mode cutoff wavelength 
-- i.e. the wavelength above which only the fundamental mode is propagated -- using the effective index approximation approach, which is a classical numerical method in guided optics \citep{Cheng1990}.\\
\indent For comparison, the bi-mode/single-mode cutoff is computed both for the {\it planar} waveguide geometry and for the {\it channel} waveguide geometry. The last value should also correspond to the cutoff wavelength for the Y-junction. 
With the parameters of Fig.~\ref{geometry}--left, we obtain for the planar waveguide a cutoff value of {\bf 8.6\,$\mu$m} and {\bf 7.7\,$\mu$m} for the TE and TM polarization of the fundamental mode, respectively. We obtain for the channel waveguide a cutoff value of {\bf 8.7\,$\mu$m} (TE) and {\bf 7.6\,$\mu$m} (TM) for the fundamental mode, respectively. Beyond these cutoff wavelengths, the guiding structure is single-mode for each respective polarization, and in particular at $\lambda$=10.6\,$\mu$m which is our operating interferometric wavelength in this work.\\
\begin{figure}[t]
\begin{minipage}{\columnwidth}	
\centering
\includegraphics[width=7.0cm,angle=180]{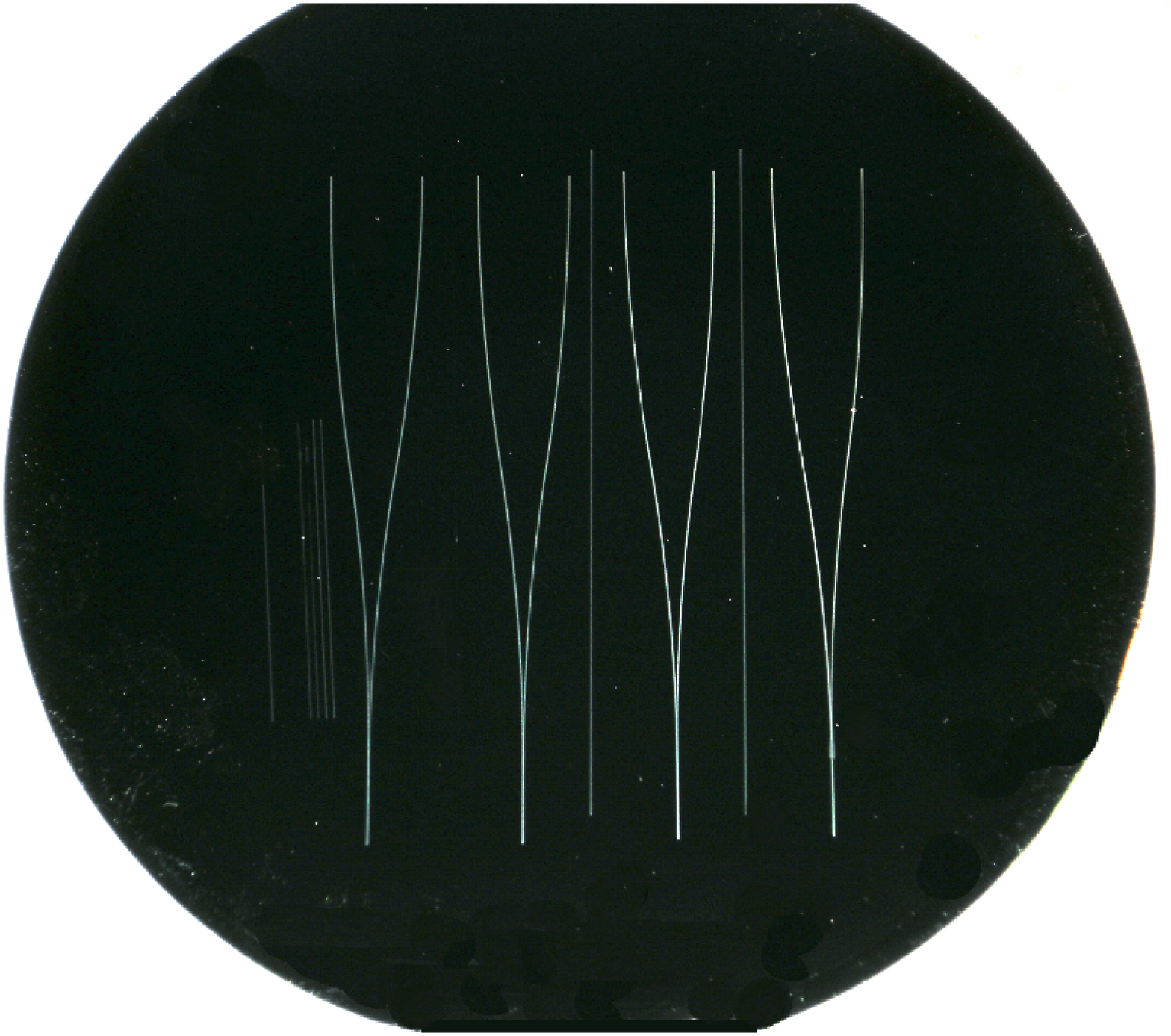} \\ \vspace{0.25cm}
\end{minipage}
\begin{minipage}{\columnwidth}
\centering
\includegraphics[width=6.0cm]{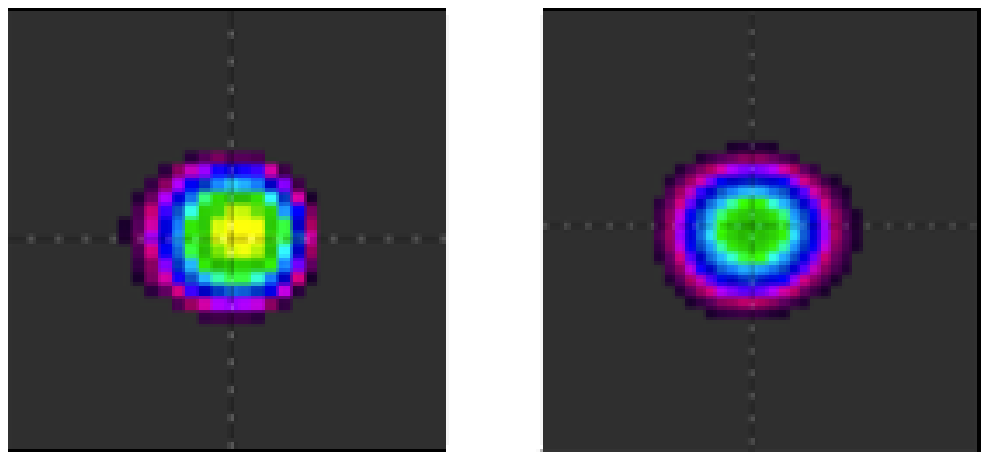}
\end{minipage}
\caption{{\it Top:} Photo of the wafer containing the Y-junctions and the channel waveguides (see text for details). {\it Down:} Qualitative snapshot views of the Y-junction outputs 
imaged at 8.5\,$\mu$m to validate the beam-splitter configuration. The field mode is unresolved by the optical system. 
}\label{component}
\end{figure}
\indent The fundamental mode 1/$e^2$ diameter is $\sim$18\,$\mu$m and $\sim$9\,$\mu$m in the horizontal and vertical directions, respectively (see Fig.~\ref{geometry}--right). This implies large divergence half-angles ($\sim$40$^{\circ}$) requiring fast coupling optics with numerical apertures of the order of $f/\#$$\sim$0.7.

\subsection{Fabrication of the Y-junction}

The manufacturing of the Y-junction device has benefited from the PNNL\footnote{PNNL: Pacific Northwest National Laboratory} expertise on the design and fabrication of chalcogenide single-mode, low-loss waveguides for the thermal infrared, which spans nominally from 8--14\,$\mu$m. 
In two previous studies, \cite{Labadie2006b} and \cite{Vigreux2007} gave evidence of the ability of Arsenic--Sulfide-Selenium (As-S-Se) structures to produce an efficient confinement of infrared radiation at $\lambda$=10.6\,$\mu$m. 
\cite{Ho2006} have demonstrated the fabrication of single-mode {\it channel} waveguides using multi-layer deposition of arsenic-based thin films followed by laser writing (Efimov et al. 2001) to provide the lateral confinement of the field. 
The new step was then to fabricate and validate more complex guiding structures useful to mid-infrared stellar interferometry.\\
\indent A chalcogenide thin film structure was produced through thermal deposition of a 3.8\,$\mu$m core layer of As$_{\rm 2}$Se$_{\rm 3}$ on a 4.5\,$\mu$m substrate layer of As$_{\rm 2}$S$_{\rm 3}$ according to deposition parameters described in \cite{Ho2006}. The As$_{\rm 2}$Se$_{\rm 3}$ film is the high-index guiding structure with $n_{\rm core}$=2.78 at 10.6\,$\mu$m while As$_{\rm 2}$S$_{\rm 3}$ forms the substrate with $n_{\rm sub}$=2.38 at 10.6\,$\mu$m. 
The lateral confinement was obtained using laser writing at $\lambda$=633\,nm and with writing power of 1\,mW, that locally photomodified the As$_{\rm 2}$Se$_{\rm 3}$ core layer, resulting in increased refractive index. 
Previous results suggests that the achieved photo-induced index difference is $\Delta n$$\simeq$0.04--0.05 under these processing conditions. 
In order to fabricate waveguides, the  As$_{\rm 2}$Se$_{\rm 3}$ thin film substrate is translated about the focal plane of the focused writing laser beam using a computer-controlled translation stage moving at a speed of 4\,mm/min. The waveguide width is slightly larger than the laser spot size of $\sim$6\,$\mu$m.\\
\indent The length of the manufactured Y-junctions is $\sim$50\,mm long with a separation of 7\,mm between the two arms. The top image of Fig.~\ref{component} shows the set of manufactured Y-junctions on the silicon wafer. This wafer also contains a set of straight waveguides written between the junctions and which are used for testing and alignment purposes. The Y-junction device was first tested for its beam-splitting function by launching light into the single input. The bottom view of Fig.~\ref{component} shows the qualitative assessment of the beam-splitting capabilities of the device at 8.5\,$\mu$m. The two Y-junction outputs were captured using an infrared microbolometer camera.


\section{Experimental setup and procedure}\label{setup}

\begin{figure}[t]
\centering
\includegraphics[width=9cm,angle=0]{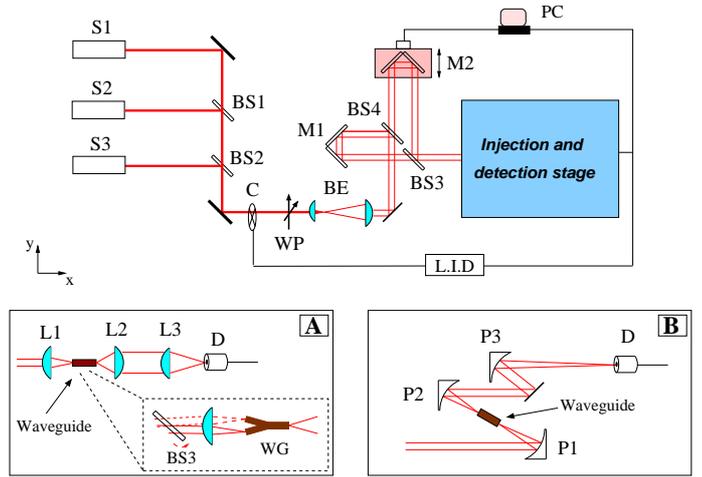}
\caption{Schematic view of the characterization testbench. {\it Top:} $S_{\rm 1}$=black body source; $S_{\rm 2}$, $S_{\rm 3}$=laser sources; $BS$=beam-splitters; $C$=chopper; $WP$=quarter-wave plate; $BE$=beam expander. {\it Inset--A:} layout for the interferometric characterization; $L$=lenses; $WG$=waveguide; $D$=detector. {\it Inset--B:} layout for the spectroscopic characterization;
$P$=off-axis parabolas.}\label{banc}
\end{figure}

Two different setups have been used in this work. One is dedicated to the interferometric characterization of the devices at $\lambda$=10.6\,$\mu$m, the second one focuses on the modal characterization of the components by Fourier-Transform spectroscopy in the 2--14\,$\mu$m range. 
The general view of the setup is presented in Fig.~\ref{banc} and we briefly describe it hereafter.\vspace{-0.4cm}
\paragraph{\bf The interferometric unit :} $S_{\rm 2}$ and $S_{\rm 3}$ are, respectively, the infrared CO$_{\rm 2}$ at $\lambda$=10.6\,$\mu$m laser and the He-Ne alignment sources. The quarter-wave plate $WP$ transforms the CO$_{\rm 2}$ flux polarization from a linear to a circular one. 
The beam-expander $BE$ produces a collimated beam of almost 1-inch diameter, which then enters the interferometer. The infrared beam is divided by a system of beam-splitters, $BS_{\rm 3}$ and $BS_{\rm4}$. 
The mirrors corner cube $M_{\rm 2}$ is mounted on a motorized translation stage and acts as the delay line to modulate the OPD. It can also be coupled to a piezoelectric motor for fine adjustment. 
In the monochromatic interferometric configuration, two options are available for waveguide coupling as depicted in the inset~A of Fig.~\ref{banc}. One option is to recombine the two beams using $BS_{\rm 3}$ before injecting into the waveguide. The second option is to slightly tilt $BS_{\rm 3}$ in order to deviate the beam coming from $M_{\rm 2}$, which produces two distinct injection spots. These are used to feed the integrated optics junction, which plays the role of the beam combiner. In this configuration, we simply use refracting optics made in Zinc Selenide (ZnSe). 
In order to be as much as compliant with the stringent numerical aperture of the current Y-junction inputs, we use an injection lens $L_{\rm 1}$ with $f$/1.1 clear numerical aperture. 
The Y-junction output flux is collected by the $L_{\rm 2}$+$L_{\rm 3}$ lenses system and focused on the  MCT\footnote{MCT or Mercury-cadmium-telluride.} detector $D$. The bench design makes it theoretically symmetric, assuming ($M_{\rm 1}$, $M_{\rm 2}$) and 
($BS_{\rm 3}$, $BS_{\rm 4}$) are ``equal'', with 3 reflections and 1 transmission per beam.\vspace{-0.5cm}
\paragraph{\bf The spectroscopic unit :} it is conceived for broadband spectral characterization of {\it channel} waveguides as depicted in inset~B of Fig.~\ref{banc}. Its configuration is very similar to the interferometric unit, but separately fed by the black-body source $S_{\rm 1}$ to cover the range from 2\,$\mu$m to 14\,$\mu$m. Hence, the light injection and collection optical system is based on off-axis parabolas ($P_{\rm 1}$, $P_{\rm 2}$, $P_{\rm 3}$) because of their achromatic nature. The broadband fringe pattern is produced by scanning the $M_{\rm 2}$ corner cube system.\vspace{-0.5cm}
\paragraph{\bf The detection unit :} The detection chain is composed of the MCT detector itself, the beam chopper $C$ to modulate the signal and a classical lock-in amplifier. The detected signal is digitally converted by the ADC card and recorded at a typical frequency of 10\,Hz or faster. The experiment takes place in open-air environment, with no active control of the OPD or the beam jitter. Our measurements can potentially be sensitive to air turbulence or mechanical micro-vibrations in the room. \vspace{0.3cm} \\
\indent In the future, we plan the development of a test-bench which allows us a direct coupling of the individual beams in the component without passing by the external beam-splitters $BS_{\rm 3}$ and $BS_{\rm 4}$, 
or to explore the viability of a fiber-link IO component (e.g. \cite{Berger1999,Haguenauer2000}) exploiting the development of single-mode chalcogenide fibers \citep{Ksendzov2007,Houizot2007}. This aspect is particularly important in view of multi-beam combination.

\section{Results and discussion}

The Y-junction is the key element of an integrated-based mid-infrared interferometer. The important physical quantities to be characterized for our prototype at this stage are the achievable interferometric contrast 
and the single-mode operation range. Other interesting parameters to be studied in the future will be the total throughput, the aging effect of the component and the impact of operation at cryogenic temperatures. 
\subsection{Temporally encoded fringes and monochromatic interferometric contrast}
\begin{figure}[t]
\centering
\includegraphics[width=8.0cm]{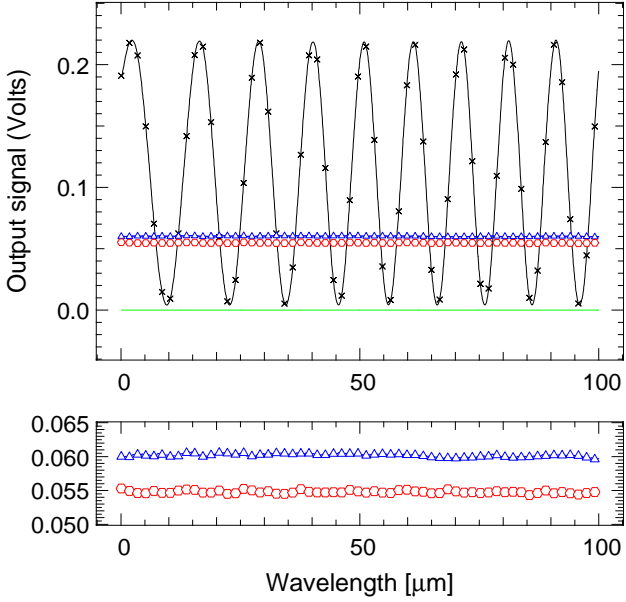}
\caption{{\it Top:} Interferometric scan obtain with the Y-junction at 10.6\,$\mu$m (black line+crosses); Photometry 1 (red curve with circles); Photometry 2 (blue curve with triangles); Bias signal (green solid line). {\it Bottom:} Zoom on the photometric channels 1 and 2 in each arm of the Y-junction over the same OPD range as for the interferometric signal.}\label{pattern}
\end{figure}
\begin{figure}[t]
\begin{minipage}{\columnwidth}
\begin{flushright}
\includegraphics[width=8.1cm]{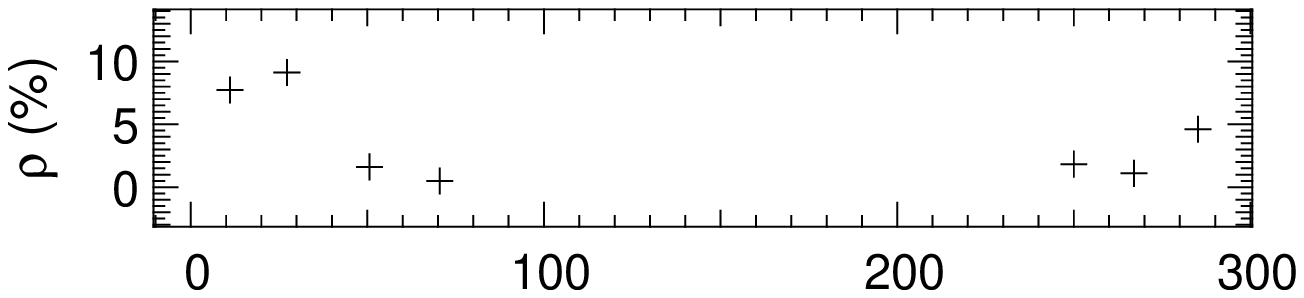}
\end{flushright}
\end{minipage}
\begin{minipage}{\columnwidth}
\begin{flushright}
\includegraphics[width=8.5cm]{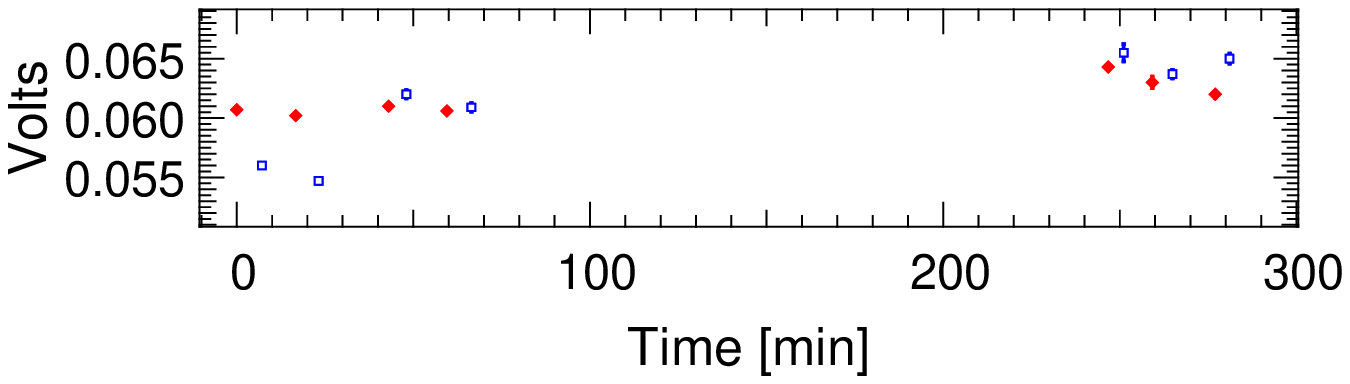}
\end{flushright}
\end{minipage}
\caption{{\it Top:} Evolution as a function of time of the photometric unbalance between the two arms of the Y-junction. 
{\it Bottom:} Evolution as a function of time of the mean and error of the photometry--1 (blue open squares) and photometry-2 (filled red diamonds) signals. In most cases, the error bar is equal or smaller than the symbol itself.}\label{rho}
\end{figure}

The Y-junction beam combiner has been characterized in the mid-infrared using the $P_{\rm 22}$ line (10.611\,$\mu$m) of a CO$_{\rm 2}$ laser. 
The setup described above permits us to inject flux in both channels, while the OPD is temporally modulated using the scanning delay line. The plots of Fig.~\ref{pattern} shows the monochromatic sinusoidal fringe pattern obtained by scanning 8 periods. The photometric signal from each of the two channels is plotted around 0.05\,V as a function of the OPD. The bias current of the detector is plotted as well, although its very low level ($\sim$10\,$\mu$V) has a negligible effect on the calibration. 
\noindent The fringe pattern displays a small chirp effect -- i.e. a broadening of the period -- from 0 to 100\,$\mu$m, resulting from known small non-linearities of the translation stage. 
However, this has no great impact on the reliability of the measurement. The bottom plot of Fig.~\ref{pattern} shows a zoom on the photometry to illustrate the small photometric unbalance between the two input channels. 
In the following, the photometry unbalance is defined by $\rho$=1-\,(I$_{\rm 1}$/I$_{\rm 2}$) with I$_{\rm 1}$ being the weakest flux, so $\rho$=0 for a perfectly balanced system. In the  case of Fig.~\ref{pattern}, the plotted curves present an unbalance of $\rho$$\simeq$\,9\%, which is then expected to have 
a negligible effect on the improvement of the interferometric contrast. Note that this last remark does not hold anymore if deep interferometric nulling is targeted, as demonstrated in \cite{Labadie2007}. However, the work presented here does not focus on optimized nulling measurement, but rather on the feasibility of an IO approach for mid-infrared interferometry. \\
\\
\indent In our specific case of a point-like source, the canonical expression for the monochromatic intensity interferogram is $I(x)=I_{\rm 1}(x)+I_{\rm 2}(x)+2\sqrt{I_{\rm 1}(x)\,I_{\rm 2}(x)}\cos(\phi(x))$ where $I$, $I_{\rm 1}$, $I_{\rm 2}$ are the interferometric, photometry\,1, photometry\,2 signals respectively. The quantity $\phi$ is the phase delay and $x$ the OPD. The photometrically calibrated interferogram $I_{\rm cor}(x)$ is simply given by 
$I_{\rm cor}(x)=\left(I(x)-I_{\rm 1}(x)-I_{\rm 2}(x)\right)/(2\sqrt{I_{\rm 1}(x)\,I_{\rm 2}(x)})$. 
$I^{\prime}(x)=1+I_{\rm cor}(x)$ is the normalized interferogram that would correspond to the observed signal if there were no photometric unbalance. 
The interferometric contrast can then be estimated in different ways, which essentially depends on the monochromatic or polychromatic conditions. We experimentally measure the raw and calibrated contrasts through $V=(I_{\rm max}-I_{\rm min})/(I_{\rm max}+I_{\rm min})$ and $V_{\rm cor}=(I^{\prime}_{\rm max}-I^{\prime}_{\rm min})/(I^{\prime}_{\rm max}+I^{\prime}_{\rm min})$. The $min$ and $max$ indices 
correspond to the maximum and minimum values for the sine signal.\\
\begin{figure}[b]
\centering
\includegraphics[width=8.0cm]{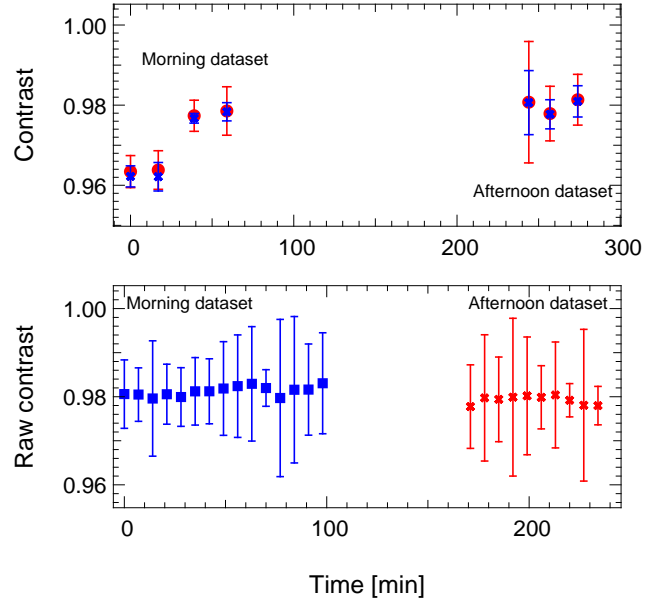}
\caption{{\it Top:} Raw (blue crosses, $V$) and calibrated (red filled circles, $V_{\rm cor}$) visibilities measured from the dataset of October 14, 2010. 
{\it Bottom:} Evolution as a function of time of the raw contrast obtained with the Y-junction on the dataset of October 6, 2010. The visibility points dispersion around the mean value is $\pm$0.001.}\label{results}
\end{figure}
\indent The plots of Fig.~\ref{results} illustrate the results of the interferometric calibration at two different dates, October 6 and October 14, 2010, respectively. The error bar associated to each visibility point is estimated through the propagation of experimental errors associated to the $I_{\rm max}$ and $I_{\rm min}$ quantities. In the top panel of Fig.~\ref{results}, we have compared the results on the raw visibilities $V$ (blue crosses) to the photometry-calibrated visibilities $V_{\rm cor}$ (red filled circles) over several hours. Each interferometric scan was interleaved with successive acquisitions of the photometry channels $I_{\rm 1}$ and $I_{\rm 2}$ under the same conditions of OPD scan. 
The individual interferometric visibilities span from 0.96 to 0.98. They appear slightly more fluctuating in the first 50\,min, then stabilizing around 0.98. 
We can notice that the photometric calibration has little or no impact on the measured interferometric contrast, which is due to the small photometric unbalance of less than 10\,\% (see Fig.~\ref{rho}) or a theoretical degradation of 0.001 of the interferometric contrast. The bottom panel of Fig.~\ref{results} shows a better sampling of the temporal variations of the raw visibilities $V$ on time scales of $\sim$10\,min. The interferometric contrast appears very stable to 0.1\% over 5\,h, with a time-averaged estimate of $V$=0.981$\pm$0.001. Here, the errors propagation method applied for each data does not appear to reflect the effective stability of the experimental visibilities, which dispersion is smaller than the individual error bars. \\
\\
\indent Our experimental measurements clearly show that the Y-junction can successfully recombine interferometrically infrared beams and deliver high and stable contrasts. 
It is also interesting for future studies to anticipate on other issues potentially affecting the interferometric performance.\\
\\
-- A first consideration is on the possible polarization mismatches that can diminish the interferometric visibility. The polarization states in the two channels are not monitored in this experiment. From \cite{Traub1988}, we know that a phase shift of 22.5$^{\circ}$ between the $s$-$p$ differences of the two beams would be sufficient to introduce a visibility drop of $V$=0.02 ($s$ and $p$ being the perpendicular and parallel components in the $xy$ plane of incidence). Although this could be a considerable loss for future high contrast interferometric measurements, 
it remains quite acceptable for traditional $V^{\rm 2}$ interferometry. Such a small phase shift could be either produced by the bench optics (despite the apparent symmetry of the interferometer) or by differential effects in the Y-junction arms. 
As an aside remark, we remind that the high dynamic range shown in \cite{Labadie2007} was obtained using a single-mode {\it and} single-polarization waveguide. Although not trivial, different practical solutions could be implemented on the bench to explore the polarization state of the combined beams before and after their coupling into the Y-junction.\\ 
\\
%
%
-- A second and more general consideration can be made on the single-mode behavior of the Y-junction. If we assume that the component is not ``perfectly'' single-mode -- i.e. that the first higher-order mode is not completely filtered out because we are too close to the cutoff for instance 
-- then we will experience residual phase errors (small misalignments, optical surfaces defects) than would not be compensated with the photometric calibration (in other words, the waveguide does not perform perfectly well the flattening of the wavefront). 
Because the single-mode behavior at a given $\lambda$ is only depending on the component properties, its design must be carefully implemented to delimit the suitable single-mode range. Eventually, a detailed analysis of the spatial distribution and propagation of the different modes in the structure as a function of the wavelength would certainly be helpful, although not critical at this stage. The question of the modal behavior is addressed in the following section. 

\subsection{Spectroscopic characterization of the modal behavior}

\begin{figure}[t]
\centering
\includegraphics[width=8.0cm]{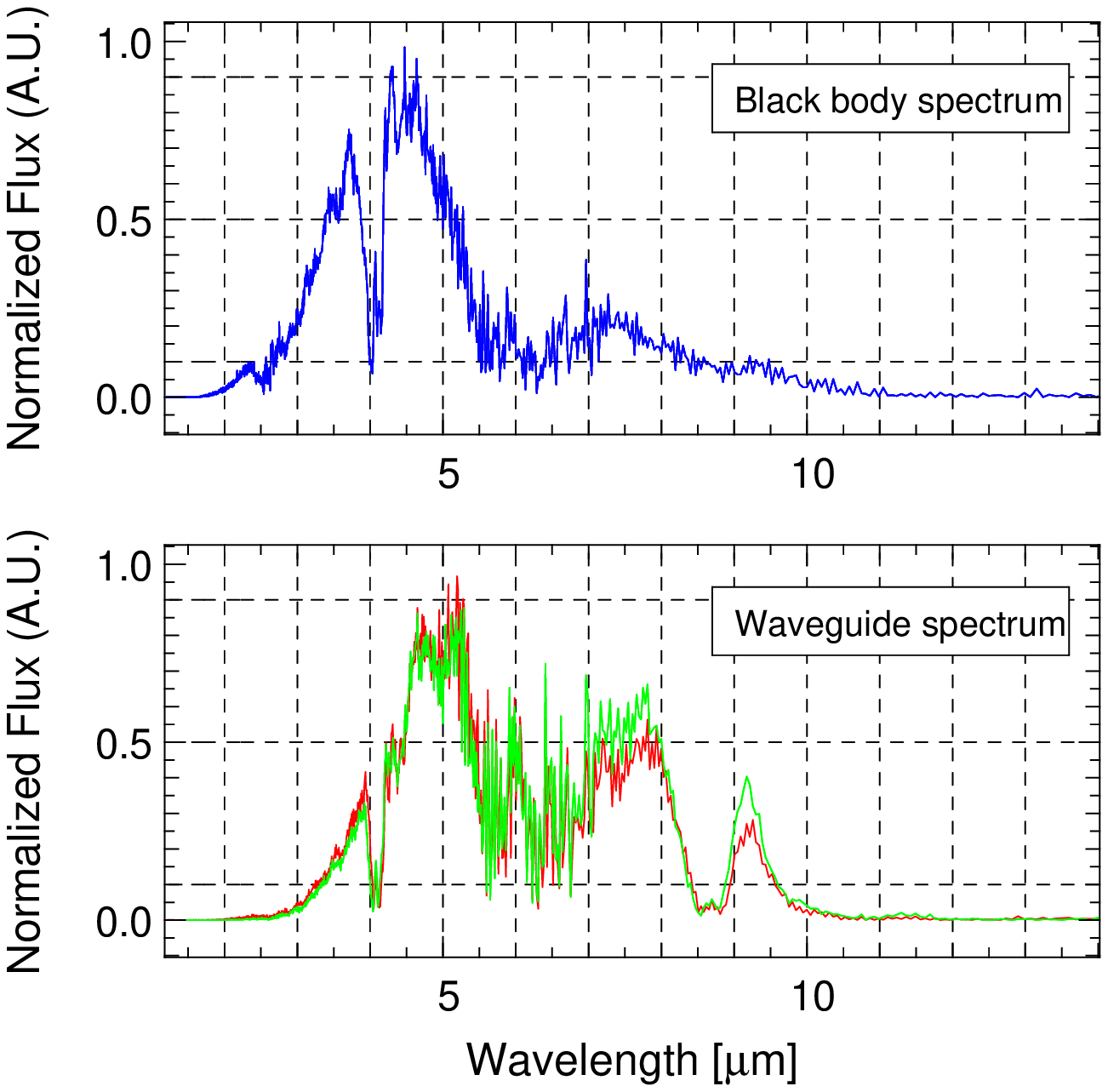}\\
\vspace{0.5cm}
\includegraphics[width=8.0cm]{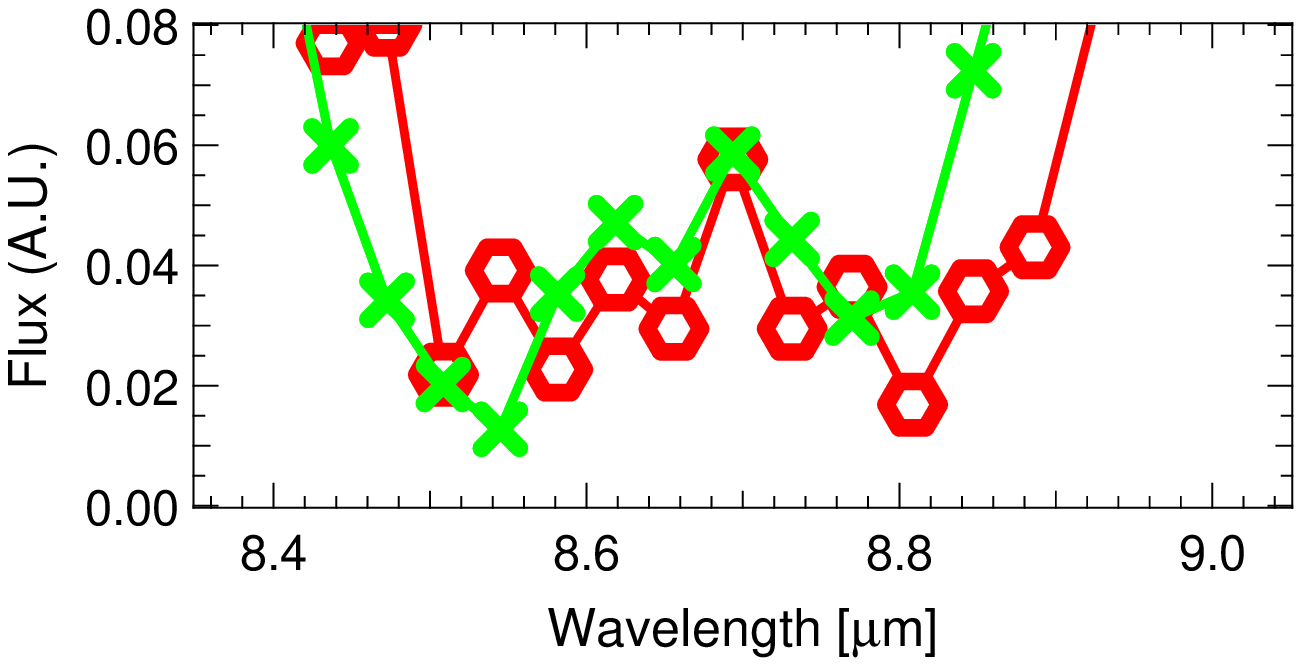}\\
\vspace{0.0cm}
\includegraphics[width=8.0cm]{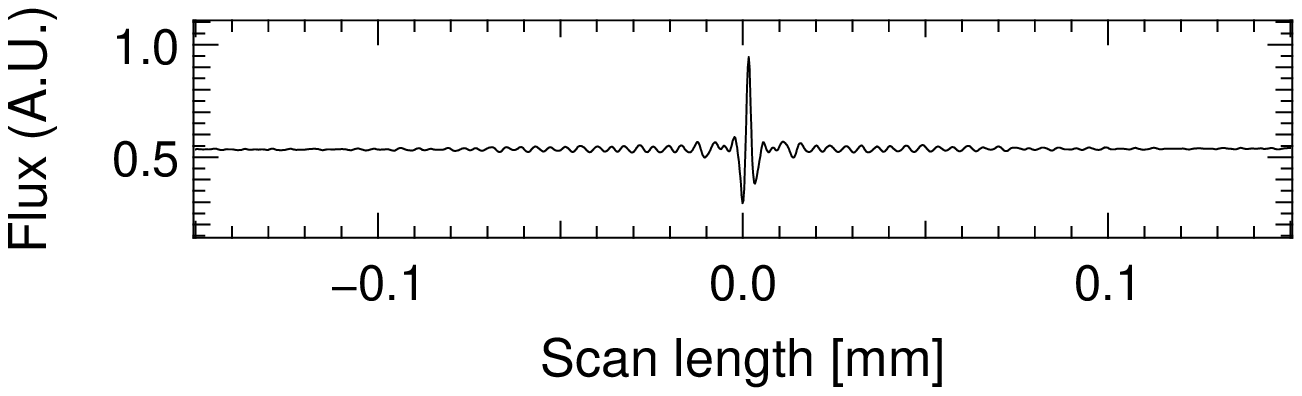}\\
\caption{
{\it Top:} Raw spectrum of the black-body source in the 2--14\,$\mu$m range. 
 transmission spectra of the raw black body 
{\it Central-top:} Two measurements (red and green curves) of the black-body source spectrum after coupling into a chalcogenide channel waveguide with opto-geometrical parameters similar to the Y-junction ones. {\it Central-bottom:} Zoom on the single-mode cutoff wavelength region. {\it Bottom:} Zoom on the typical white light interferogram obtained with the FTS.
}\label{spectro}
\end{figure}

As explained in Sect.~\ref{Intro}, the single-mode feature of the Y-junction is essential for an efficient spatial filtering of the incoming wavefront. In order to take advantage of the wavefront filtering capabilities of the component, 
it is therefore necessary to assess the spectral region where solely one mode is guided through the waveguide. By definition, the component presents several cutoff wavelengths, which defines different modal regimes (e.g. tri-mode, bi-mode, single-mode). These different regions can be evidenced through the {\it transmitted power method} \citep{Lang1994}, a technique routinely used for the characterization of single-mode fibers, and implemented as well in the field of astronomical instrumentation \citep{Laurent2002}. The principle of this method is based on the idea that the energy coupled into the waveguide at a given wavelength is distributed among the different propagation modes supported by the waveguide. As the wavelength varies from shorter to longer values, the number of supported modes decreases by one at each cutoff wavelength, and this can be traced in a transmission spectrum through the identification of an abrupt dip in the guided flux intensity \citep{Grille2009}.\\
\indent This technique has been used here to characterize the modal behavior of the channel waveguides visible in Fig.~\ref{component} next to the Y-junction, which were manufactured following the same procedure. We verified the single-mode behavior of the component at 10\,$\mu$m 
by doing Fourier Transform spectroscopy between 2 and 14\,$\mu$m with the setup described in Sect.~\ref{setup}. 
The bottom plot of Fig.~\ref{spectro} shows a zoom on the central white light fringe, which is acquired over a total OPD length of 2\,mm resulting in a spectral resolution $\Delta\nu$=5\,cm$^{\rm -1}$. 
We have produced one reference spectrum of the black body source to be compared with the waveguide spectra, which both are presented in the top panels of Fig.~\ref{spectro}.
\\
\indent The blue curve in the top panel is  the raw emission spectrum of the black-body source acquired in the sensitivity range of the MCT detector. It includes the transmission by the lab atmosphere and the testbench components. Typical features can be observed such as the CO$_{\rm 2}$ absorption line around 4.2\,$\mu$m or the water vapor absorption lines after 2.5\,$\mu$m and between 5\,$\mu$m and 7\,$\mu$m. In the top-mid panel, the green and red curves are, under same experimental conditions, two different measurements of the previous spectrum after coupling light into the chalcogenide channel waveguide. The black body and waveguide spectra have been normalized to their peak intensity: indeed, because of the detection parameters (i.e. gain of the lock-in amplifier) that changed between the two measurements, the two curves cannot be compared directly. Here the goal is limited to the identification of {\it relative} variations in the spectrum resulting from the waveguide characteristics. \\ 
\indent The ``waveguide'' spectra include transmission effects related to the coupling input and output coupling efficiency, to the transparency of the As$_{\rm 2}$Se$_{\rm 3}$ glass (i.e. propagation losses), and to the modal behavior of the waveguide which is precisely the sought effect. On this last point, as this effect results from the loss of successive propagation modes from shorter to longer wavelengths, the amplitude of the transmission dip will be stronger as we go towards the lower order cutoff wavelengths. Hence, we have chosen to focus here only on the search for the deepest bi-mode/single-mode transition since higher order shallower jumps at shorter wavelengths would be more difficult to identify among the several features present in the spectrum. Therefore, the {\it transmitted power method} comes as a supporting technique to experimentally trace, in a restricted spectral range, the modal characteristics inferred theoretically, 
rather than a method for blind-searches of cutoff frequencies.\\
\indent The waveguide plots of Fig.~\ref{spectro} reveal a clear intensity drop from $\sim$8.0\,$\mu$m down to $\sim$8.5\,$\mu$m in both measurements of the waveguide spectrum. 
A small double-dip feature is observable at $\sim$8.5--8.6\,$\mu$m and $\sim$8.8\,$\mu$m before the curve rises up. 
A large intrinsic absorption feature is very unlikely considering the homogeneous transparency of these chalcogenide glasses in this wavelength range \citep{Savage1965,Moynihan1975}. 
Then, a possible interpretation is that we observe the bi-mode/single-mode transition of the waveguide with a measured cutoff around 8.5\,$\mu$m. 
The rise-up of the curve afterwards 8.8\,$\mu$m corresponds to the well-known effect of energy transfer from first order to the fundamental mode following the cutoff. 
The double-dip feature observed at 8.6\,$\mu$m and 8.8\,$\mu$m in the waveguide spectrum may actually correspond to the close signatures of the planar and the channel waveguide, theoretically expected at 8.6\,$\mu$m and 8.7\,$\mu$m. Indeed, because the injection spot of the FTS setup is currently considerably larger than the channel waveguide cross-section ($\sim$30\,$\mu$m vs. $\sim$5\,$\mu$m), a non-negligible fraction of the input energy is likely coupled into the surrounding planar waveguide, rather than solely in the channel waveguide, and collected back onto the detector. 
In this hypothesis, we find for the TE polarization a good agreement between the experimental and the theoretical values derived in Sect.~\ref{modal-behav}, which are intrinsically very close. The 0.1\,$\mu$m shift in the cutoff wavelength between theory and experience is within the range of uncertainty introduced by the manufacturing process on the index difference $\Delta n$ and on the physical size of the waveguide, which ultimately define the modal behavior. Being able later to reduce the spot size, keeping at the same time a good signal-to-noise ratio, would certainly help to minimize this effect.  
For the TM polarization, the cutoff wavelength is expected by design at 7.7\,$\mu$m, i.e. in a spectral region where we do not benefit from a good SNR due to the water absorption band. This suggests that the TM cutoff signature may be hidden in the quite noisy region of the spectrum. 
Eventually, in the solid hypothesis of an observed cutoff wavelength and in spite of the small uncertainty raised above, this result confirms experimentally the single-mode behavior of the manufactured structure around $\lambda$=10\,$\mu$m.

\section{Conclusion}

-- We conducted a first laboratory study at 10\,$\mu$m of the interferometric capabilities of a single-mode integrated-optics beam combiner in the context of instrumentation activities for multi-aperture IR interferometry. The Y-junction, manufactured by laser-writing, is composed of chalcogenide glasses, which transparency is useful for integrated-optics beam combination in the L\,(3.2\,$\mu$m), L$^{\prime}$\,(3.8\,$\mu$m), M\,(4.5\,$\mu$m) and N\,(10\,$\mu$m) bands. 
\vspace{0.15cm}\\
-- We have demonstrated a high fringe contrast of 0.981$\pm$0.001 at $\lambda$=10.6\,$\mu$m, with high repeatability. The fringe contrast shows to remain very stable over a significant time period of 5\,h, and even between two consecutive weeks. The component is also quite stable with regards to photometry variations (Fig.~\ref{rho}-bottom) possibly resulting from time dependent laser drifts, mechanical instabilities of the motorized delay light impacting light coupling into the waveguides, or thermal instabilities of the laboratory environment. We also verified and experimentally confirmed the single-mode range of the device beyond 8.5\,$\mu$m. \vspace{0.15cm}\\
-- In this study, we have also experienced the importance of a proper numerical aperture optimization between the coupling optics and the waveguide structure. 
Large throughput losses occur from both the numerical aperture mismatch and the Fresnel losses at each facet. The magnitude of these losses was not evaluated in this study. 
We are confident this can be solved by a well-coordinated design of the component and of the interface optics at the telescope. However, we already know from previous studies \citep{Ho2006} that the intrinsic propagation losses of this type of device are as low as $\sim$0.5\,dB/cm. This is promising for the immediate future, where wide-band interferometric fringes constitute the next objective. \vspace{0.15cm}\\
\indent Although additional research in this area is required, this study can be seen as a first step towards the design, fabrication, and optimization of future instruments that exploit the benefits of astro-photonics \citep{Bland-Hawthorn2009} at mid-infrared wavelengths. This work also shows that we are not so far from a new important milestone that could be the on-sky validation of this elegant approach on an operating interferometer. On the longer term, the development of balloon-borne prototype interferometer at mid-infrared wavelengths comparable to what is already being achieved in the far-IR \citep{Shibai2010} could probably benefit from such advances.


\begin{acknowledgements}
LL is funded by the Spanish MICINN under the Consolider-Ingenio 2010 Program grant CSD2006-00070:First Science with the GTC (www.iac.es/consolider-ingenio-gtc). This work was also supported by the U.S. Department of Energy, Office of Nonproliferation Research and Development (NA-22). Pacific Northwest National Laboratory is operated for the U.S. Department of Energy by Battelle Memorial Institute under Contract No.~DE-AC05-76RLO1830.
\end{acknowledgements}

\bibliographystyle{aa}

\bibliography{paper_gc.bib}

\end{document}